\documentclass[preprint, 12pt]{elsart}

\usepackage{graphicx}

\usepackage{amssymb}

\journal{Nuclear Physics A}

\begin{document}

\begin{frontmatter}



\title{Calculations of $K^-$ nuclear quasi-bound states based on chiral 
meson-baryon amplitudes.}


\author{Daniel Gazda and Ji\v{r}\'{\i} Mare\v{s}}

\address{Nuclear Physics Institute, 250 68 \v{R}e\v{z}, Czech Republic}

\begin{abstract}
In-medium ${\bar K}N$ scattering amplitudes developed within a new 
chirally motivated coupled-channel model due to Ciepl\'{y} and Smejkal that fits 
the recent SIDDHARTA kaonic hydrogen $1s$ level shift and width are used 
to construct $K^-$ nuclear potentials for calculations of $K^-$ nuclear quasi-bound 
states. The strong energy and density dependence of scattering amplitudes 
at and near threshold leads to $K^-$ potential depths 
$-{\rm Re}V_K \approx 80 -120$~MeV. Self-consistent calculations of all  
$K^-$ nuclear quasi-bound states, including excited states, are reported. 
Model dependence, polarization effects, the role of $p$-wave interactions, and 
two-nucleon $K^-NN\rightarrow YN$ absorption modes are discussed. 
The $K^-$ absorption widths $\Gamma_K$ are comparable or even larger than the 
corresponding binding energies $B_K$ for \emph{all} $K^-$ nuclear quasi-bound states, 
exceeding considerably the level spacing. This discourages search for $K^-$ nuclear 
quasi-bound states in any but lightest nuclear systems.

\end{abstract}


\end{frontmatter}

\section{Introduction}
\label{intro}

The study of the interaction of antikaons with baryonic systems, such as kaonic atoms, 
$K^-$ nuclear clusters or dense strange kaonic matter, is an interesting issue 
with far-reaching consequences, e.g. for heavy-ion collisions and astrophysics.   
The closely related problem of $K^-$ nuclear quasi-bound states is far from being 
settled, despite much theoretical and experimental effort in the last decade~\cite{npa08,hypx}.    

Our first studies of $K^-$ quasi-bound states in nuclear many-body systems based 
on an extended relativistic mean field (RMF) model were focused on  the widths expected 
for $K^-$ quasi-bound states \cite{mfg06}. The subject of multi-$K^-$ 
(hyper)nuclei was studied in Refs.~\cite{gfgm07,gfgm08,gfgm09} with the aim to  
explore whether kaon condensation could occur in strong-interaction self-bound 
baryonic matter. In the RMF formulation, the energy independent real $K^-$ nuclear 
potential was supplemented by a phenomenological imaginary potential fitted to 
kaonic atom data, with energy dependence that accounted for the reduced 
phase space available for in-medium $K^-$ absorption, including 2-nucleon 
absorption modes. 
             
Weise and H\"{a}rtle performed calculations of $K^-$ nuclear states in $^{16}$O and $^{208}$Pb 
using chiral-model ${\bar K}N$ amplitudes within a local density approximation~\cite{wh08}. 

This paper reports on our latest calculations of $K^-$ nuclear quasi-bound states 
within a chirally motivated meson-baryon coupled-channel separable interaction model \cite{cs10}.  
We apply a self-consistent scheme for constructing $K^-$ nuclear potentials from 
subthreshold in-medium ${\bar K}N$ scattering amplitudes which was introduced 
in Refs. \cite{cfggm1, cfggm2}. This time, the ${\bar K}N$ amplitudes are constructed using 
a recent in-medium coupled channel model NLO30 \cite{cs11}  that reproduces all available low energy 
${\bar K}N$ observables, including 
the latest $1s$ level shift and width in the  $K^-$ hydrogen atom from 
the SIDDHARTA experiment \cite{sid11}. We demonstrate the crucial role of the strong 
energy and density dependencies of the $K^- N$ scattering amplitudes, leading  to 
deep $K^-$ nuclear potentials for various considered versions of 
in-medium modifications of the scattering amplitudes. Using several  
versions of the chirally motivated coupled-channel model, we demonstrate the model 
dependence of our calculations. Moreover, we discuss the effects of $p$-wave interactions  
and the $K^-NN \rightarrow YN$ absorption modes. Finally, we present 
binding energies and widths of {\it all} $K^-$ quasi-bound states -- 
including excited states -- in selected nuclei.  

The paper is organized as follows. In Section 2, we briefly describe the model 
and underlying self-consistent scheme for constructing $K^-$ nucleus potentials 
from in-medium subthreshold ${\bar K}N$ scattering amplitudes. 
In Section 3, we present results of our calculations of $K^-$ quasi-bound 
states in various nuclei across the periodic table. 
Conclusions are summarized in section 4. 

\section{Model}
\label{sec:1}

In this section, we briefly outline the methodology which forms the framework 
of our calculations of $K^-$ nuclear quasi-bound states. We concentrate only 
on basic ingrediences of the model since the details can be found in 
Refs. \cite{cfggm1, cfggm2, cs11}.  

The $K^-$ nuclear quasi-bound states are determined by self-consistent 
solution of the in-medium Klein--Gordon equation:
\begin{equation}
\label{eq:kg}
\left[ \nabla^2+\tilde{\omega}_K^2-m_k^2 - \Pi_K(\omega_K,\vec{p}_K,\rho) \right] \phi = 0
,
\end{equation}
where $\tilde{\omega}_K$ is complex energy of antikaon containing the Coulomb
interaction $V_C$ introduced by minimal substitution:
\begin{equation}
\tilde{\omega}_K = \omega_K - \mathrm{i} \Gamma_K/2 - V_C,  
\end{equation}
with $\Gamma_K$ being the width of $K^-$ nuclear state of energy $\omega_K =  m_K - B_K$, where  
$B_K$ is the binding energy of antikaon.
The self-energy operator $\Pi_K = 2({\rm Re}\, \omega_K)V_K$ is constructed 
in a ``$t \rho$'' form with the
amplitude calculated in a chirally motivated coupled-channel approach:

\begin{equation}
\label{eq:seo}
\Pi_K(\omega_K,\vec{p}_K,\rho) = - 4\pi\frac{\sqrt{s}}{m_N}\left[
F_{K^-p} (\sqrt{s},\vec{p},\rho) \rho_p
+F_{K^-n} (\sqrt{s},\vec{p},\rho) \rho_n
\right]
,
\end{equation}
where $F_{K^-p(n)}$ is the $K^-$--proton (neutron) in-medium scattering amplitude 
in a separable form, $\vec p$ is the relative $K^-N$ momentum, and 
$\sqrt{s}$ is the $K^-N$ total energy. 
The realistic proton $\rho_p$ and neutron $\rho_n$ density
distributions in the core nuclei are taken from relativistic mean-field 
nuclear-structure calculations.
The $K^-N$ scattering amplitudes $F_{K^-N}$ are constructed within a chirally motivated 
coupled-channel separable interaction model \cite{cs10}. 
In this work, we applied the latest version NLO30 \cite{cs11} which 
reproduces the recent $K^-$ hydrogen $1s$ level shift and width from the 
SIDDHARTA experiment \cite{sid11}. For the sake of comparison, we applied as well 
the TW1 model \cite{cfggm2,cs11} fitted also to the SIDDHARTA data,  
and the older CS30 model \cite{cs10}. It is to be noted that in the TW1 (NLO30, CS30) model, the effective
separable meson-baryon potentials are constructed to match the equivalent amplitudes derived 
from the chiral effective Lagrangian at leading (next-to-leading) order, respectively.

When the elementary $K^-N$ system is embedded in the nuclear medium of density $\rho$, 
one has to consider in-medium modifications of the scattering amplitude, in particular 
 Pauli blocking in the intermediate states (this in-medium version is marked `no SE') 
\cite{wkw96}. 
The other version (marked `+SE') adds self-consistently meson and baryon self-energies 
generated by the interaction of hadrons with the nuclear medium \cite{lu98,ro00}. 

\begin{figure}[hbt] 
\begin{center} 
\includegraphics[width=0.45\textwidth]{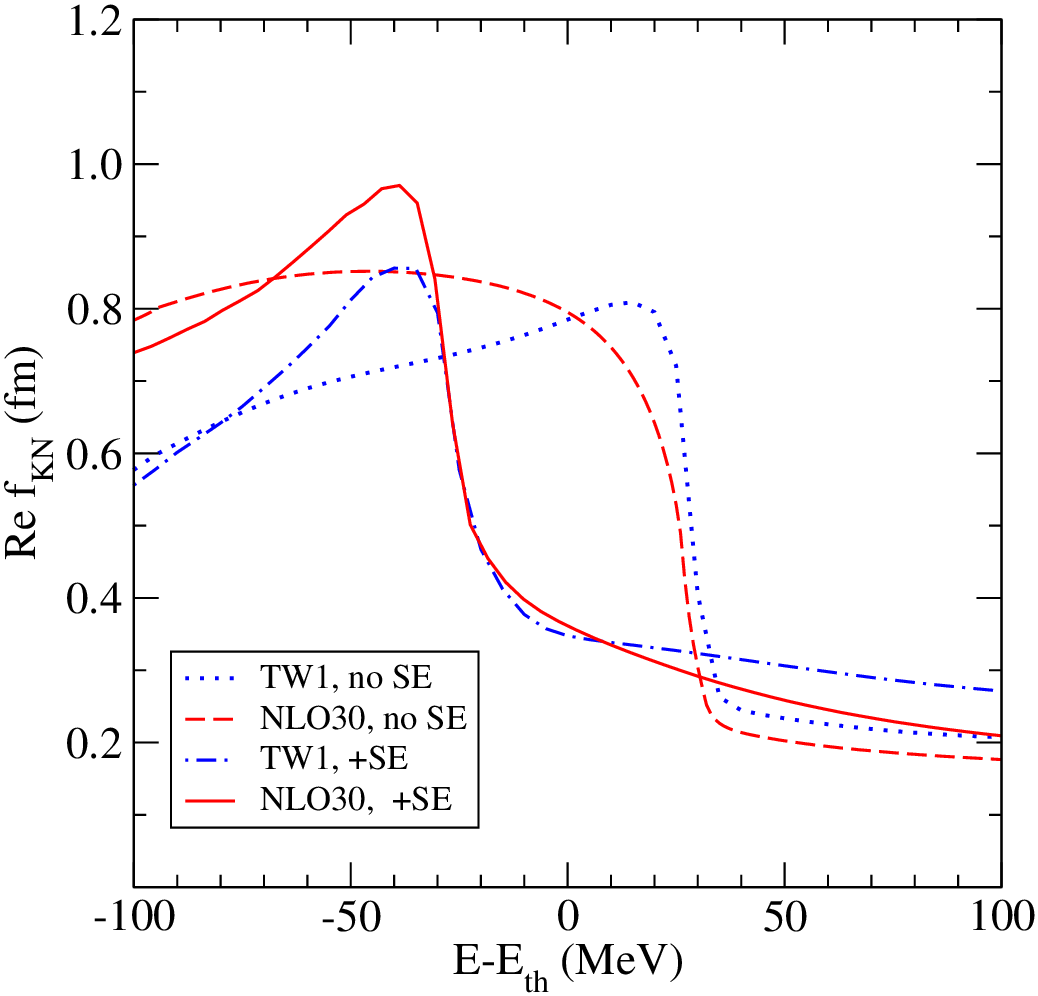} 
\hspace{3mm} 
\includegraphics[width=0.45\textwidth]{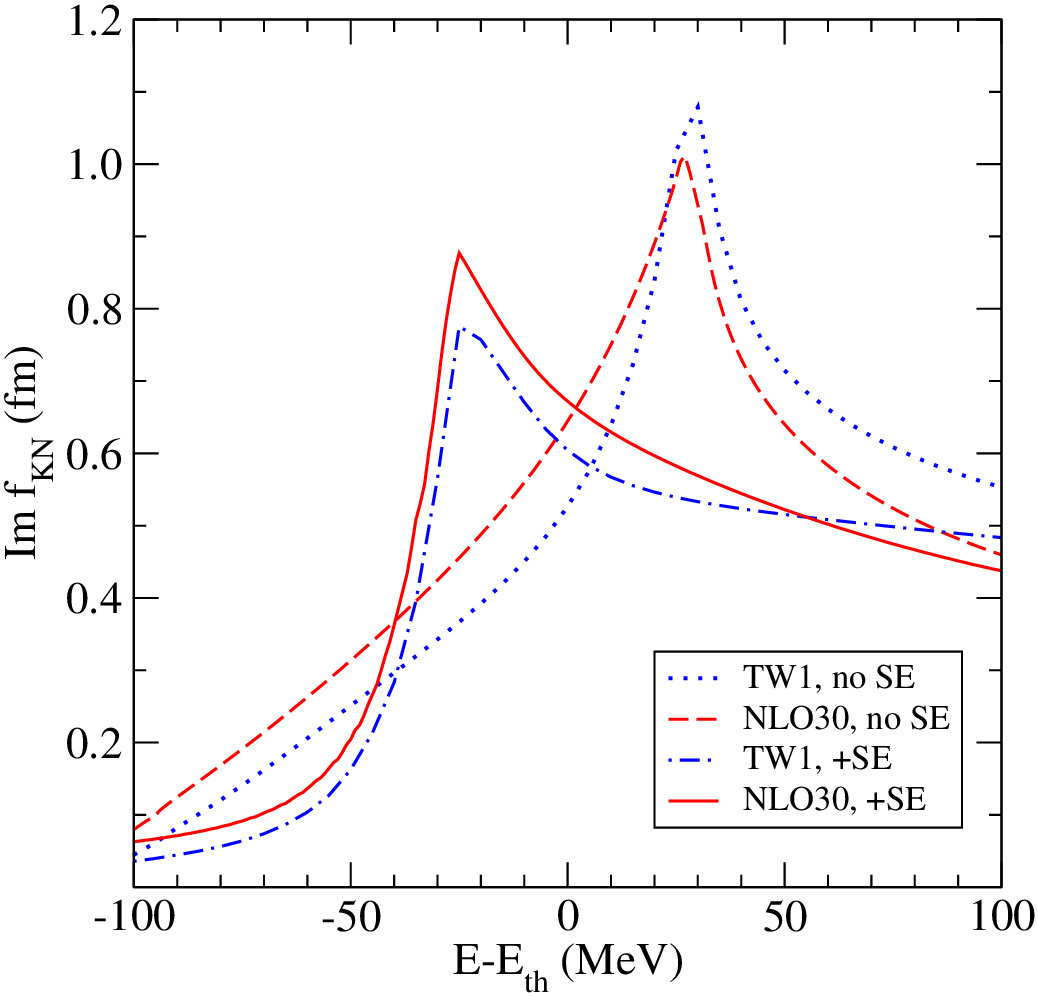} 
\caption{Energy dependence of the c.m.\ reduced amplitude $f_{K^-N}$ in TW1 and NLO30 models (left: real part, right: imaginary part).   
Dotted and dashed lines: Pauli blocked amplitude (`no SE') for $\rho_0 = 0.17$ fm$^{-3}$; solid and dot-dashed lines: 
including also hadron self-energies (`+SE') at $\rho_0$.} 
\end{center} 
\end{figure} 
Figure 1 illustrates a typical resonance-shape energy dependence of the in-medium reduced 
\footnote{$F_{K^-N} = g(p)f_{K^-N}g(p')$, see Ref. \cite{cfggm2}} scattering amplitudes  
$
f_{K^-N}(\sqrt{s},\rho)=1/2 [f_{K^-p}(\sqrt{s},\rho) + f_{K^-n}(\sqrt{s},\rho)]\; 
$
for nuclear matter density $\rho_0=0.17$~fm$^{-3}$, 
corresponding to the interaction of the $K^-$ meson with symmetric nuclear matter.
The amplitudes were calculated within the TW1 and NLO30 models and for each of the models,  
two in-medium versions, `no SE' and `+SE', are shown for comparison. 
The pronounced energy dependence of the scattering amplitude appears crucial in
the self-consistent calculations of kaonic nuclear states. In particular, the real part 
of the `+SE' amplitudes in both models changes from weak attraction at and above threshold to 
strong attraction at $\sim 30$ MeV below threshold. As a result, the `+SE' and `no SE' 
amplitudes become close to each other at energies relevant for self-consistent calculations 
of kaonic nuclei. 
While both models give similar imaginary parts Im$f_{K^-N}$ for each in-medium version, they 
differ considerably ($\approx 20\%$) in real parts of the scattering amplitudes below threshold. 
This indicates the extent of the model dependence of the calculations of $K^-$ nuclear quasi-bound states.  

The scattering amplitude $F_{K^-N}$ in Eq.\ (\ref{eq:seo}) is a function of
$K^-N$ c.m.\ energy $\sqrt{s}$ and relative momentum $\vec{p}$. For nuclear
bound-state applications it is necessary to transform the two-body $K^-N$
arguments into the $\bar{K}$--nuclear c.m.\ frame. For the relative momentum 
$\vec{p}$ we have 
\begin{equation}
\vec{p} = \xi_N\vec{p}_K-\xi_K\vec{p}_N,\quad \xi_{N(K)}=m_N(K)/(m_N+m_K),
\end{equation}
which upon averaging over angles yields $p^2$ of the form
\begin{equation}
p^2 = \xi_N \xi_K
\left(
2 m_K \frac{p_N^2}{2m_N}+2m_N\frac{p_K^2}{2m_K}
\right).
\end{equation}
Similarly, expanding near threshold energy  $E_{\rm th} = m_K + m_p$, 
$\sqrt{s}$ assumes the form 
\begin{equation}
\sqrt{s} \approx 
E_{\rm th} -B_K -V_C - B_N - \xi_N \frac{p_N^2}{2m_N}
- \xi_K \frac{p_K^2}{2m_K}
,
\end{equation}
where $B_N$ stands for the binding energy of nucleon, and
the last two terms represent corrections due to the kinetic energies of nucleon
and antikaon. The nucleon kinetic energy is approximated in the
Fermi gas model $p^2_N/(2m_N) = 23 (\rho / \rho_0)^{2/3}$~MeV and the kinetic energy of
antikaon is obtained by means of local density approximation $p^2_K/(2m_N) =
-B_K -\mathrm{Re}\,\mathcal{V}_K$, where ${\mathcal V}_K=V_K +V_C$.
 This finally leads to 
\begin{eqnarray}
p^2\approx & \xi_N\xi_K\left[ 2 m_K\, 23 \left(\rho/\rho_0
\right)^{2/3}-2m_N(B_K+\mathrm{Re}\,\mathcal{V}_K(\rho)) \right] \;\;\;\mathrm{(in\; MeV),} 
\\
\label{eq:s}
\sqrt{s} \approx & E_{\rm th} - B_N - \xi_N B_K - 15.1 
\left( \frac{\rho}{\rho_0}\right)^{2/3} + \xi_K {\rm Re}{\mathcal V}_K(\rho)
 \;\;\;\;\mathrm{(in\; MeV).} 
\end{eqnarray}
We note that the $K^-$ potential $V_K$ and the $K^-$ binding energy $B_K$ appear 
as arguments in the expression 
for $\sqrt{s}$ , which in turn serves as an argument for the self-energy $\Pi_K$,    
and thus for $V_K$. This suggest a self-consistency scheme in terms of both 
$V_K$ and $B_K$ for solving the Klein-Gordon equation (1). 
  
It is to be stressed that the present chiral model of $K^-$--nucleus
interaction does not account for the absorption of $K^-$ mesons in the nuclear medium
through nonpionic conversion modes on two nucleons $K^-NN\rightarrow YN$
($Y=\Lambda ,\Sigma$). To estimate the contribution of two-nucleon absorption
processes to the
decay widths of $K^-$ nuclear states we introduced phenomenological term
into the $K^-$ self-energy:
\begin{equation}
{\rm Im}\, \Pi_K^{(2N)} = 0.2 f_{YN}(B_K) W_0 \rho^2
,
\end{equation}
where $W_0$ was fixed by kaonic atom data analysis and $f_{YN}(B_K)$ is
kinematical suppression factor taking into account reduced phase space available
for decay products of $K^-$ nuclear bound states \cite{mfg06}.

The present chiral model also does not address the $p$-wave part of the
$K^-$--nucleus interaction. Since $p$-waves may play an important role for
tightly bound $K^-$ nuclear systems \cite{wg05} we studied their contribution  
by adding the self-energy term:
\begin{equation}
\Pi_K^{(p)} = -4 \pi \frac{m_N}{\sqrt{s}} {\vec \nabla} C_{K^-N}(\sqrt{s}) \rho \cdot
{\vec \nabla}
,
\end{equation}
with $C_{K^-N}$ $p$-wave amplitude constrained by the $\Sigma (1385)$ 
resonance phenomenology and parametrized following Ref.\ \cite{wh08}. 

\section{Results and discussion}
\label{sec:2}

We adopted the above methodology to calculations  of quasi-bound $K^-$ states in 
selected nuclei across the 
periodic table. In most cases, we solved KG equation self-consistently in a static 
approximation.  For the sake of comparison, we performed also fully dynamical 
calculations upon taking into account the polarization of the nuclear core by the 
strongly bound antikaon.   

\begin{figure}[b!] 
\begin{center} 
\includegraphics[width=0.47\textwidth]{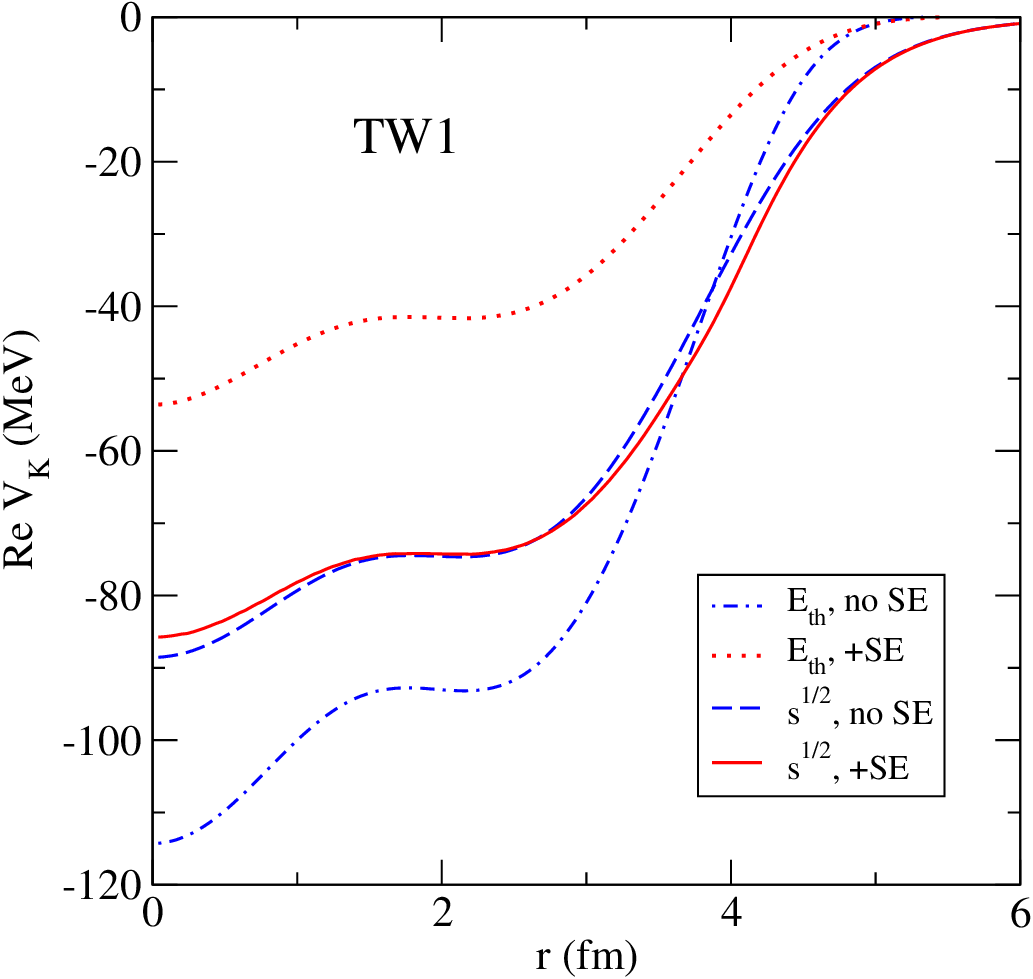} 
\hspace{3mm} 
\includegraphics[width=0.47\textwidth]{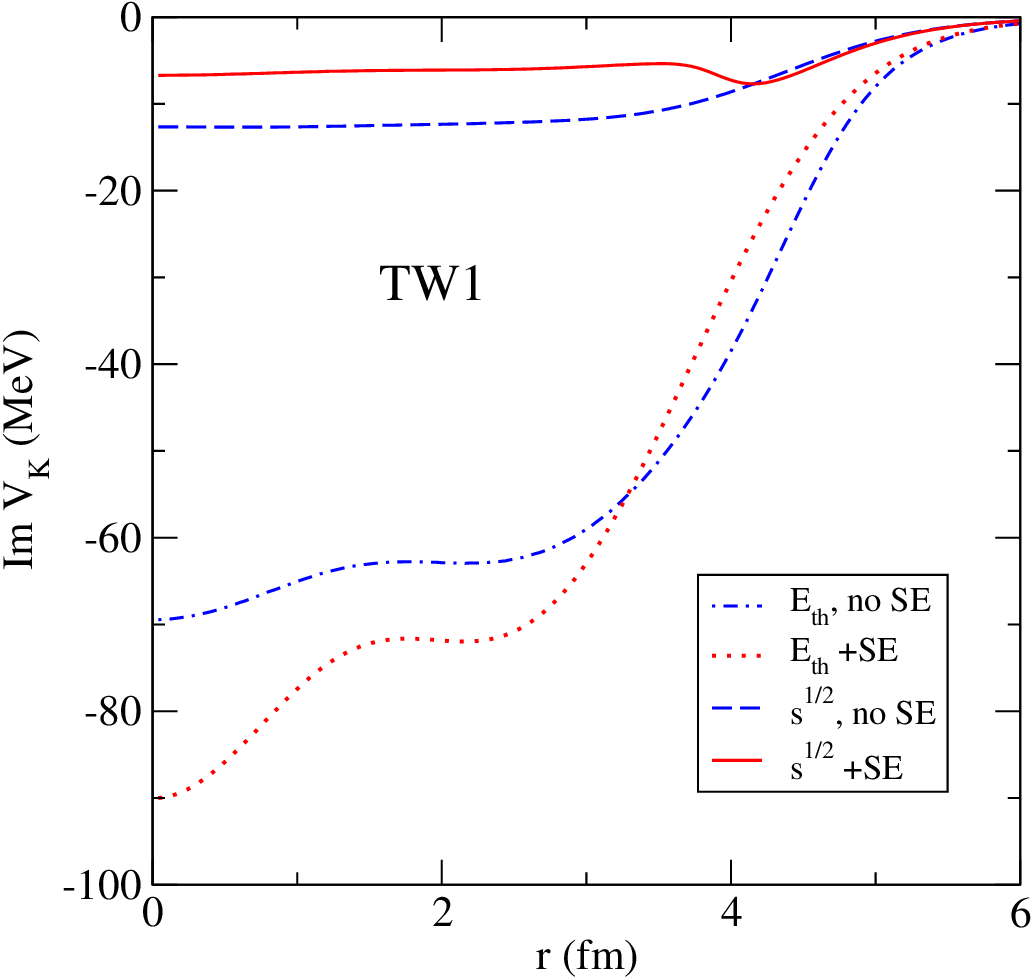} 

\includegraphics[width=0.47\textwidth]{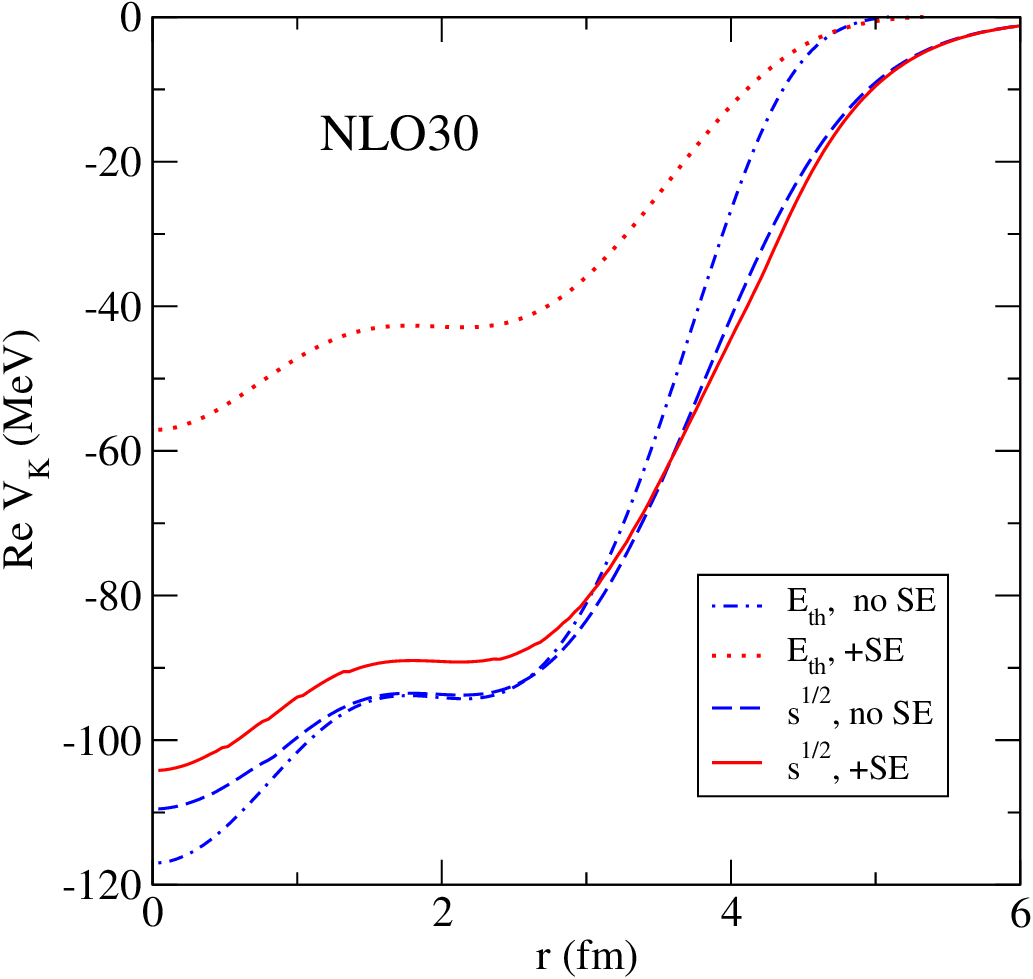} 
\hspace{3mm} 
\includegraphics[width=0.47\textwidth]{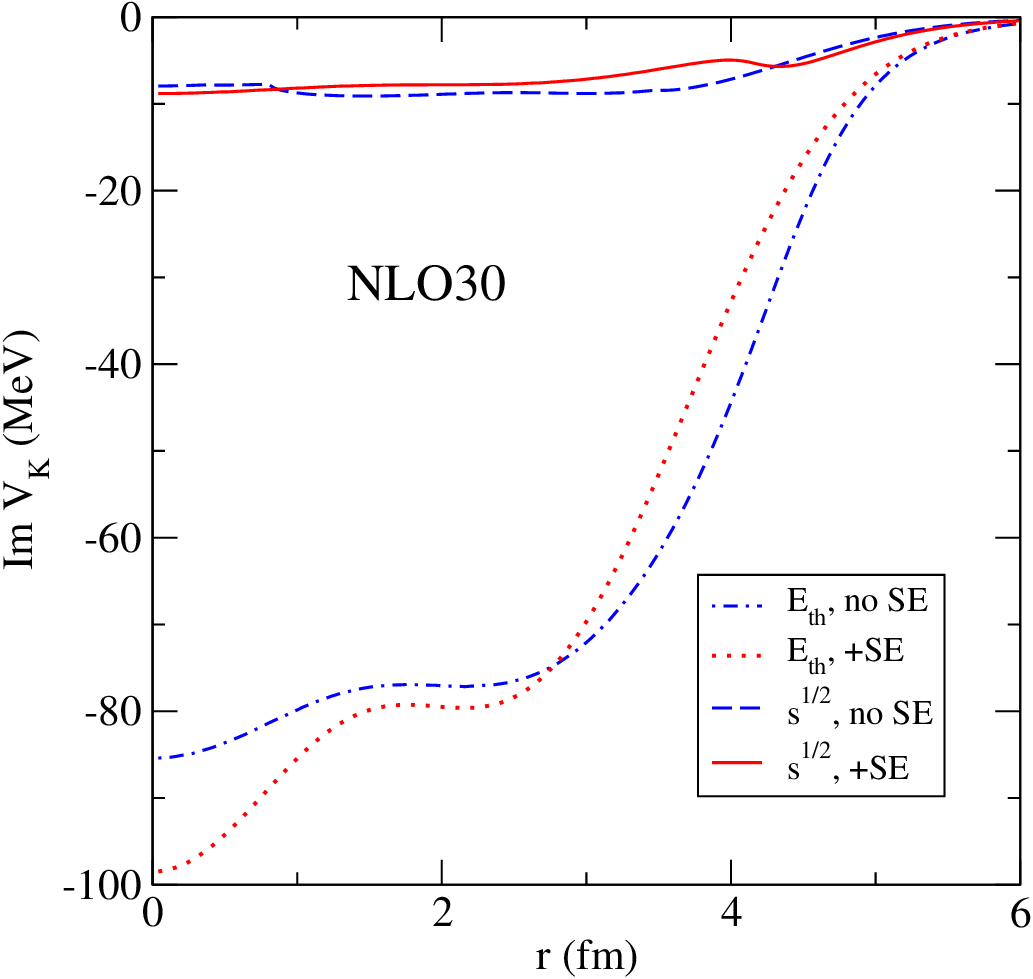} 
\caption{$K^-$ nuclear potentials in Ca (left: real part, 
right: imaginary part),  calculated with static RMF nuclear densities and chiral 
amplitudes (upper panels: TW1, lower panels: NLO30)  at threshold 
(`$E_{\rm th}$') and with $\sqrt{s}$, in both in-medium 
versions: only Pauli blocking (`no SE') and including also hadron self-energies (`+SE').} 
\end{center} 
\end{figure} 
 
Figure~2 shows the $K^-$ nuclear potentials for $1s$ state in Ca at threshold (`$E_{\rm th}$') and for $\sqrt{s}$
(Eq.\ (\ref{eq:s})) calculated self-consistently with respect to Re$V_K$ and $B_K$, 
for TW1 amplitudes (upper panels) and NLO30 amplitudes
(lower panels). 
The relevant values of $B_K$ in Ca and other nuclei are presented in the following figures.  
It is seen that the subthreshold extrapolation $\sqrt{s}$ is crucial for the depth 
of $V_K$, calculated using the TW1 amplitude in both `no SE' and `+ SE' in-medium version. 
While at threshold the  depth of Re$V_K$ in the `+SE' case is about half of the depth 
in the `no SE' case, for $\sqrt{s}$ both in-medium versions give a similar depth 
-Re$V_K(\rho_0)\approx 85-90$~MeV. 
In the case of the NLO30 model, the situation is rather different. While the self-consistent
calculation is still important for the depth of Re$V_K$, evaluated using the `+SE' in-medium
modification, in the `no SE' case, the depths of Re$V_K$ evaluated at threshold and at 
$\sqrt{s}$ are close to each other. This weak energy dependence of Re$V_K$ has 
its origin in rather weak energy dependence of the `no SE' NLO30 scattering amplitude below threshold, 
as shown in Fig.~1.  The NLO30 model yields $\sim 20$~MeV deeper Re$V_K$ than the TW1 model, 
as could be anticipated from Fig.\ 1. 
The imaginary parts of $V_K$ and consequently the widths which represent only $K^-N \rightarrow YN$ 
decays, are considerably reduced in the self-consistent calculations 
of the subthreshold amplitudes owing to the proximity of the $\pi\Sigma$ threshold (see both right
panels of Fig.\ 2). 

\begin{figure}[t!]
\begin{center}
  \includegraphics[width=0.72\textwidth]{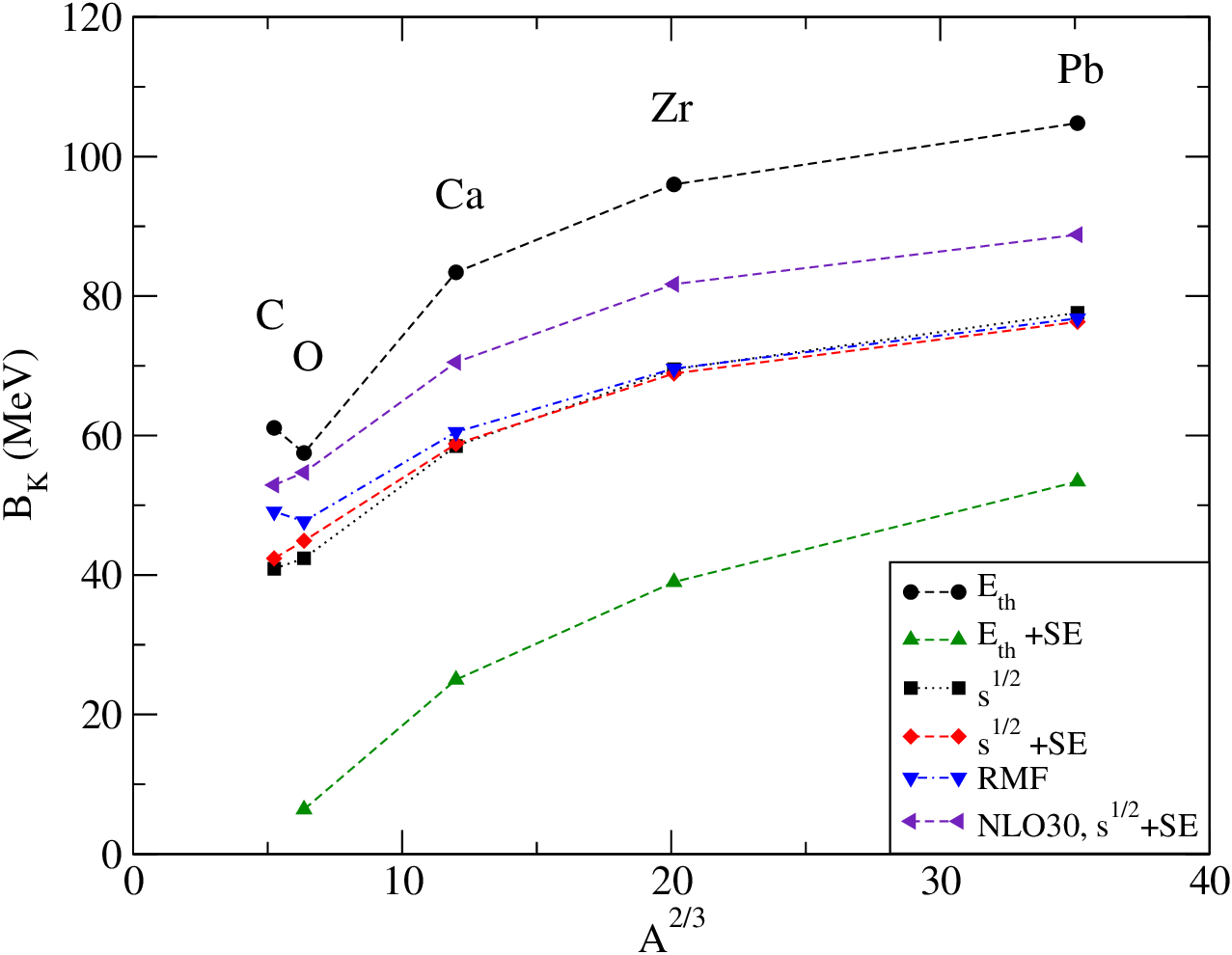}
\end{center}
\caption{Binding energies $B_K$ of $1s$ $K^-$ nuclear quasi-bound states in several nuclei, 
calculated using static RMF nuclear densities and TW1 chiral amplitudes at threshold 
(`$E_{\rm th}$') and with $\sqrt{s}$, in both in-medium versions: 
only Pauli blocking (`no SE') and including also hadron self-energies (`+SE'). Results of static RMF 
calculations, with a $K^-$ nuclear interaction mediated by vector mesons only, as well as for 
NLO30 chiral `+ SE' amplitudes are shown for comparison.}
\end{figure}
\begin{figure}[thb]
\begin{center}
  \includegraphics[width=0.72\textwidth]{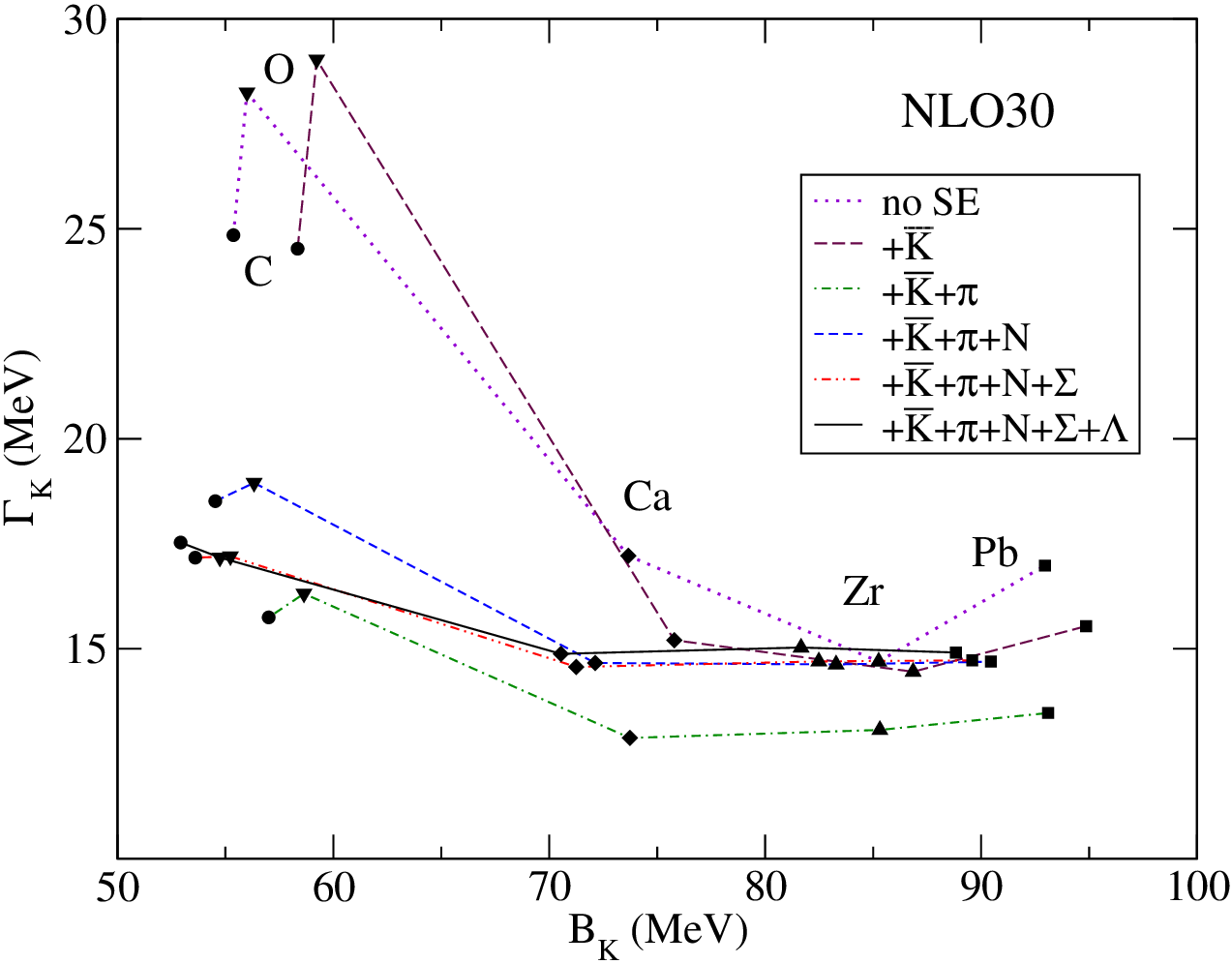}
\end{center}
\caption{Binding energies $B_K$ and widths $\Gamma_K$ of $1s$ $K^-$ nuclear quasi-bound states,
 calculated self-consistently with static RMF nuclear densities and in-medium NLO30 amplitudes 
with various hadron self-energies considered in intermediate states (see the legend). 
$K^-NN \rightarrow YN$ decay modes are not
included.  
}
\end{figure}

The peculiar role of the energy dependence of the $K^-N$ scattering amplitudes is illustrated for the
TW1 model in Fig. 3. Binding energies $B_K$ of $1s$ $K^-$ nuclear quasi-bound states 
obtained by solving Eq.\ (\ref{eq:kg}) self-consistently (denoted `$s^{1/2}$') for several nuclei  
are compared with $B_K$ calculated using threshold amplitudes ($E_{\rm th}$). 
It is worth noting that the self-consistent calculations of $B_K$ using `no SE' and `+SE' 
in-medium amplitudes give very similar results. These $B_K$ values are remarkably close to 
those calculated within a static RMF approach, when the $K^-$-nucleus interaction is mediated 
exclusively by vector mesons with purely vector SU(3) F-type couplings (denoted `RMF')~\cite{gfgm07}. 
For comparison, we present also binding energies $B_K$ calculated self-consistently using 
the NLO30 `+SE' amplitudes which are more than 10 MeV larger than the corresponding binding 
energies calculated using the TW1 amplitudes.    

Figure 4 shows the effect of particular hadron self-energies in the intermediate states on 
the $K^-$ binding energies $B_K$ and widths $\Gamma_K$, calculated self-consistently within 
the NLO30 model. The widths $\Gamma_K$ are plotted as function of $B_K$ for various options of
implementation of the self-energies, as indicated in the legend of Fig. 4. The binding
energies $B_K$ in all nuclei under consideration differ in all cases by less than 5 MeV and the 
effect of self-energies seems to be A independent. Implementation of pion 
self-energies leads to a sizable reduction of the widths $\Gamma_K$ in lighter nuclei (C, O). 
On the other hand, the role of hyperon self-energies seems to be marginal.     

It is to be noted that the calculated widths shown in the figure represent only 
$K^-N \rightarrow \pi Y$ decays, accounted for by the coupled-channel chiral model.   
When phenomenological energy dependent imaginary $\rho^2$ terms are added self-consistently 
to simulate two-nucleon $K^-NN \rightarrow YN$  absorption modes and their available phase 
space~\cite{mfg06}, the resulting widths of order $\Gamma_K \approx 50$~MeV become comparable 
in light nuclei to the binding energies $B_K$.

Figure 5 illustrates the model dependence of the $K^-$ nuclear state calculations by presenting 
$B_K$ and $\Gamma_K$ of $1s$ $K^-$ states in selected nuclei calculated self-consistently 
using various chiral-model scattering amplitudes -- TW1, CS30, and NLO30.    
The CS30 and NLO30 models yield larger binding energies and lower widths of the nuclear $K^-$ 
quasi-bound states than the TW1 model. The differences between the chiral models are up to 
20 MeV in the binding energies and up to 10 MeV in the widths. The model dependence of $B_K$  
and $\Gamma_K$ is thus more pronounced than the effects of self-energies in the self-consistent 
calculations of kaonic nuclei. 

\begin{figure}[t!]
\begin{center}
  \includegraphics[width=0.7\textwidth]{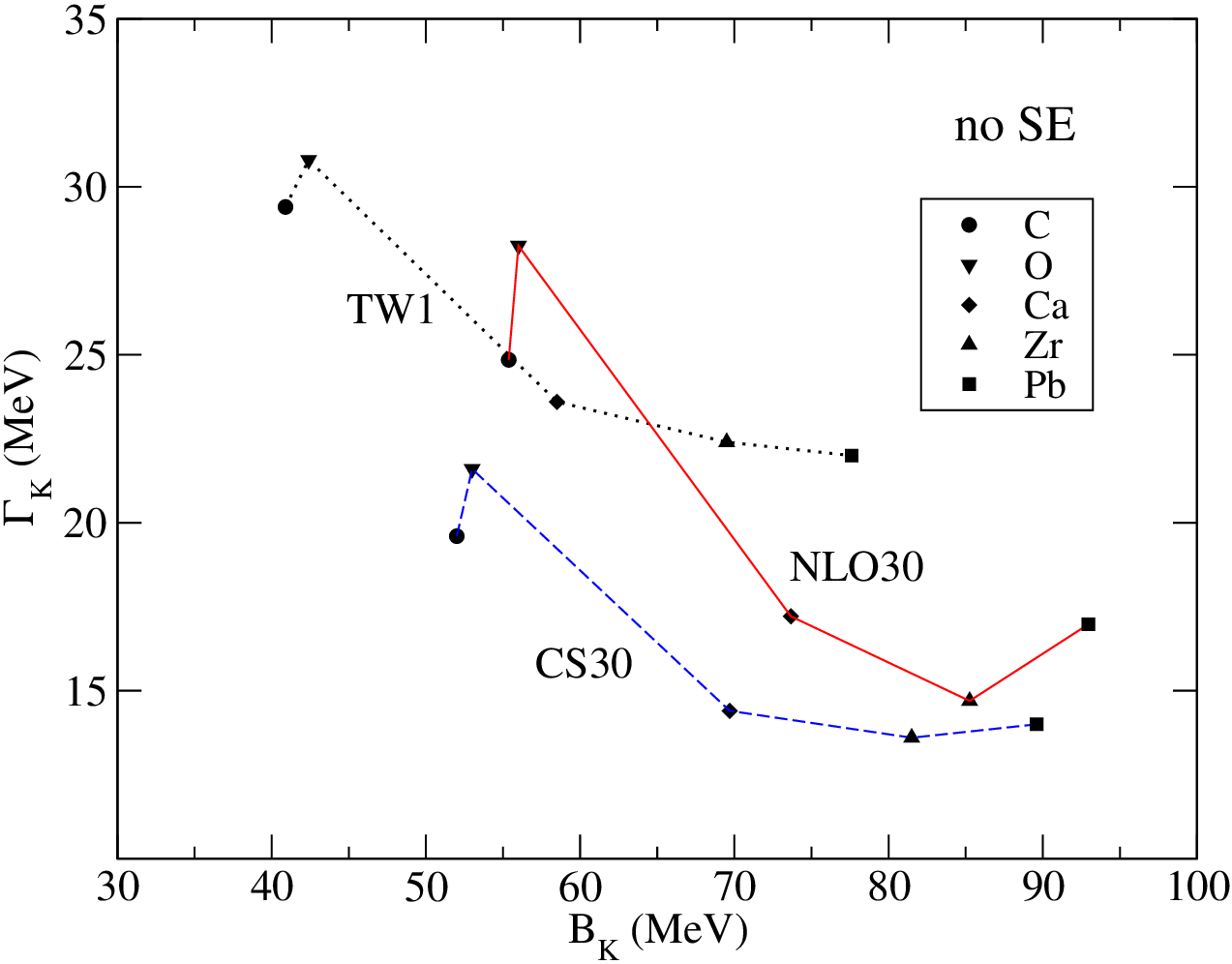}
\end{center}
\caption{Binding energies $B_K$ and widths $\Gamma_K$ of $1s$ $K^-$ nuclear 
quasi-bound states, calculated self-consistently with static RMF densities and the `no SE' 
TW1, CS30 and NLO30 scattering amplitudes.   
$K^-NN \rightarrow YN$ decay modes are not included. 
}
\end{figure}

\begin{figure}[b!] 
\begin{center} 
\includegraphics[width=0.48\textwidth]{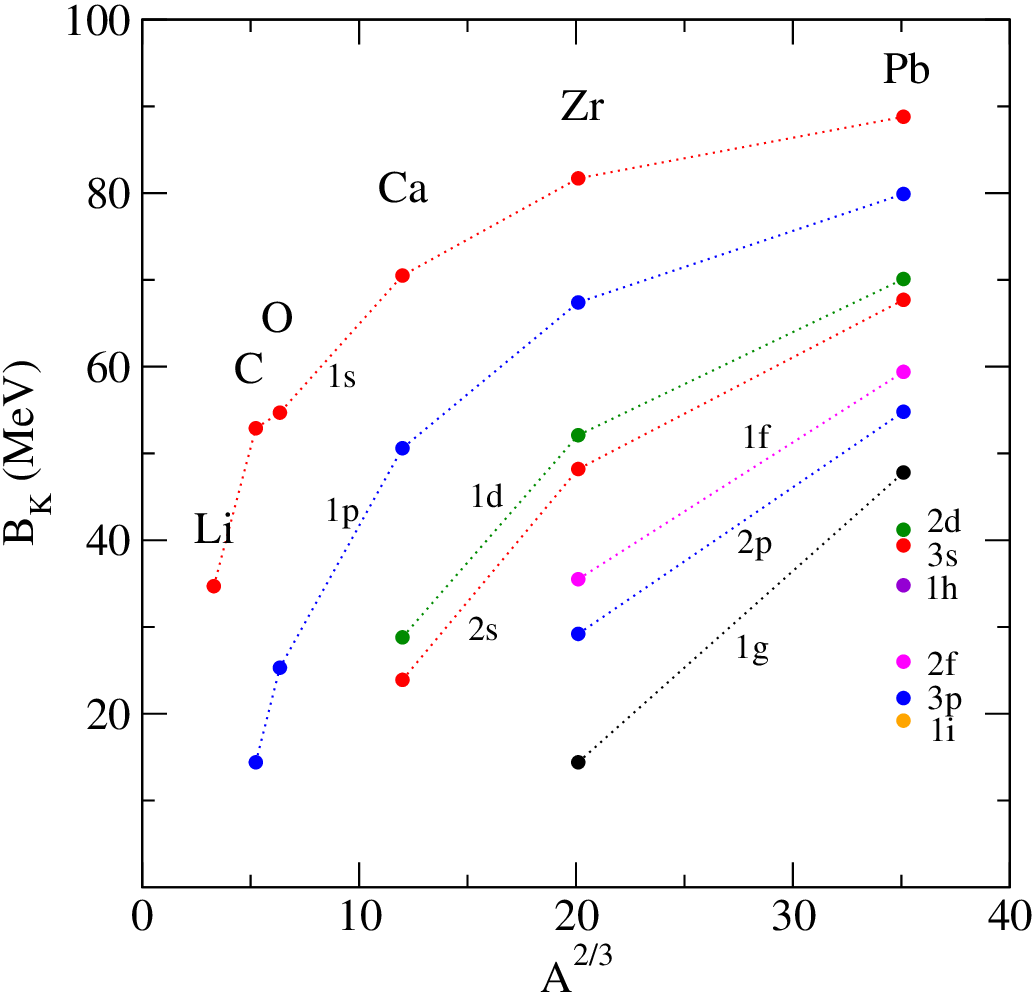} 
\hspace{3mm} 
\includegraphics[width=0.48\textwidth]{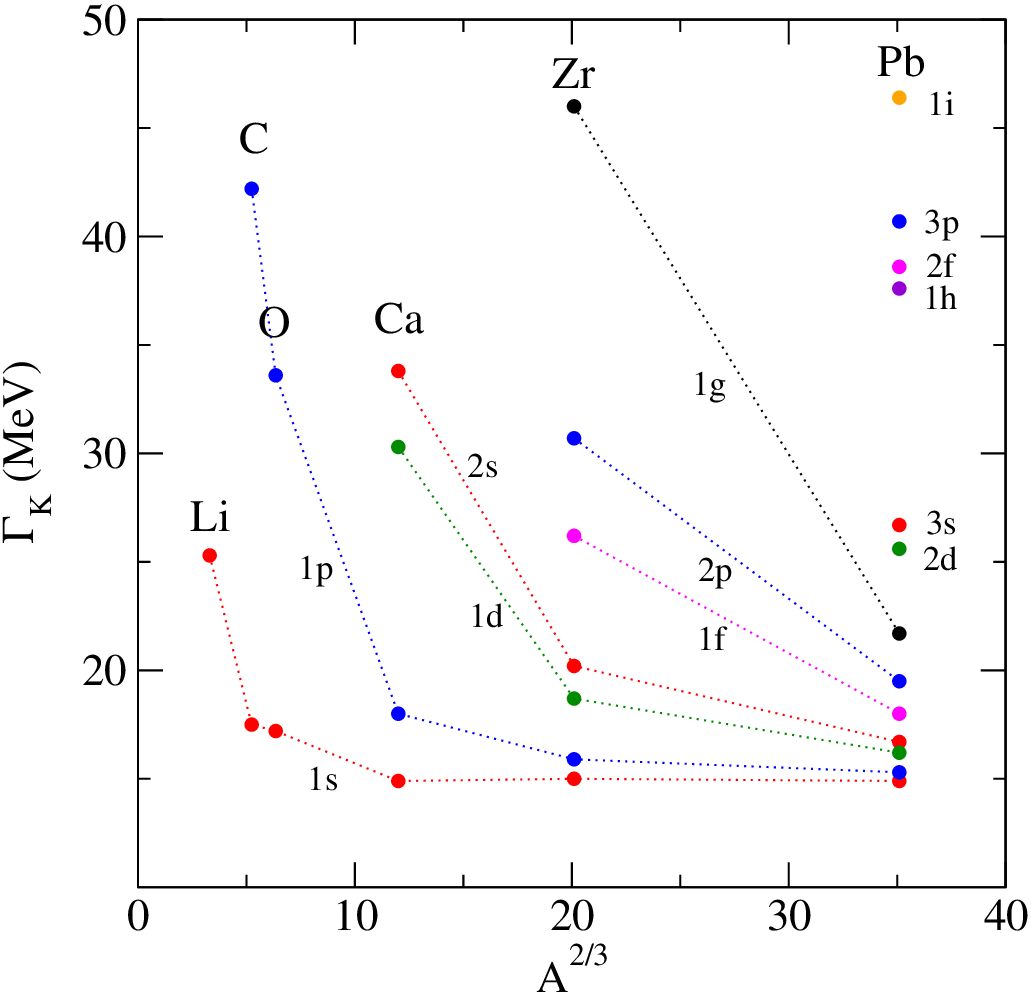} 
\caption{Binding energies $B_K$ (left panel) and widths $\Gamma_K$ (right panel) of $K^-$ 
quasi-bound states in selected nuclei, calculated self-consistently with static RMF densities and the 
`+ SE' NLO30 scattering amplitudes. $K^-NN \rightarrow YN$ decay modes are not included.}  
\end{center} 
\end{figure} 

Figure 6 shows binding energies and widths of $K^-$ quasi-bound states -- including excited states -- in selected nuclei calculated by applying self-consistently the 
subthreshold extrapolations $\sqrt{s}$ of Eq.\ (\ref{eq:s}) to the NLO30 `+SE' amplitudes. Clearly, the widths 
of higher excited states are appreciable, exceeding the corresponding level spacing, even in the case  
when $2N$ absorption is not considered. This leads to considerable overlap of the $K^-$ quasi-bound 
states even in the lightest nuclei (except Li, where only the $1s$ $K^-$ quasi-bound state exists). 
It is to be stressed that the $2N$ absorption which was not considered here, adds additional sizable  
contribution to the widths, particularly of low lying states. Such large widths thus inevitably obscure 
experimental study of the $K^-$ nuclear quasi-bound states in heavier nuclei.        

\begin{table}[b!]
\begin{center}
\caption{
Binding energies $B_K$ and widths $\Gamma_K$ (in MeV) of the 
$K^-$ nuclear quasi-bound states in Ca, calculated self-consistently using NLO30 `+SE' 
amplitudes. Dynamical and static RMF schemes are compared in the first two blocks. 
Results of a static RMF scheme including $p$-wave amplitudes is shown in the third block, 
and $K^-NN \rightarrow YN$ decay modes are included in the last block (`+2N abs.').
}
\begin{tabular}{|c|cc|cc|cc|cc|}
\hline
 & \multicolumn{2}{|c|}{\hspace*{3mm}dynamical\hspace*{3mm}} & \multicolumn{2}{|c|}
{\hspace*{8mm}static\hspace*{8mm}} & 
\multicolumn{2}{|c|}{stat. + $p$ wave} & \multicolumn{2}{|c|}{stat. + $2N$ abs.} \\  
&  $B_K$  & $\Gamma_K$ &  $B_K$  & $\Gamma_K$ &  $B_K$  & $\Gamma_K$ &  $B_K$  & $\Gamma_K$  \\
\hline
$1s$  & 72.3 & 14.8 & 70.5 & 14.9 & 73.0 & 14.8 & 68.9 & 58.9 \\[2mm]
$1p$  & 52.8 & 17.7 & 50.6 & 18.0 & 53.1 & 17.9 & 49.2 & 53.6 \\[2mm]
$1d$  & 30.5 & 29.2 & 28.8 & 30.3 & 32.1 & 29.3 & 27.7 & 59.7 \\[2mm]
$2s$  & 24.6 & 30.9 & 23.9 & 33.8 & 26.3 & 34.2 & 21.6 & 67.1 \\[2mm]
\hline
\end{tabular}
\label{tab:B}
\end{center}
\end{table}

Table 1 shows, as representative examples, binding energies $B_K$ and widths $\Gamma_K$ 
of $K^-$ nuclear quasi-bound states in Ca, calculated self-consistently
using the NLO30 `+SE' scattering amplitudes. The results of fully dynamical RMF calculations
which take into account the polarization of the nuclear core by the strongly bound $K^-$ meson,  
are compared with the static RMF scheme in the first 2 blocks. 
The dynamical calculations give, in general, higher binding energies $B_K$ and smaller widths $\Gamma_K$. 
As could be anticipated, the polarization effect is A dependent: while it increases $B_K$ by $\sim 6$~MeV 
in Li, it is less than 2~MeV  in Ca (shown in Table 1), and in Pb the difference between static and 
dynamical calculations is less than 0.5~MeV. 
Effects of adding a $p$-wave $K^-N$ 
interaction assigned to the $\Sigma(1385)$ resonance  are demonstrated within the static RMF scheme 
in the third block of the table. 
The $p$-wave 
interaction increases the $K^-$ binding energy only by few MeV, being more pronounced 
in light nuclei where surface effects are relatively more important. Nevertheless, even in Pb the 
$p$-wave interaction increases $B_K$ by $\sim 2$~MeV. 
Finally, phenomenological energy dependent imaginary $\rho^2$ terms are 
included to simulate $2N$ absorption processes $K^-NN \rightarrow YN$.   
Whereas the $K^-$ binding energies $B_K$ decrease only slightly, the absorption widths 
$\Gamma_K > 50$~MeV become comparable to the binding energies $B_K$ or even much larger in the case 
of higher $K^-$ nuclear quasi-bound states.

\section{Conclusions}
\label{sec:3}
We performed extensive study of the $K^-$ nuclear quasi-bound states within a chirally motivated 
meson-baryon coupled-channel separable interaction model. 
We considered two in-medium versions of the $K^-N$ scattering amplitudes: the `no SE' version 
which takes into account only Pauli blocking in the intermediate states, and the 
`+SE' version which adds self-consistently hadron self-energies. In addition, we used several 
 versions of the model to explore model dependence of our calculations.  
In this contribution, we demonstrate on few selected examples main results of the calculations 
with the aim to assess the role of various ingredients of the approach that influence binding 
energies and widths of the $K^-$ nuclear states. 
Energy dependence of the in-medium scattering amplitudes, particularly in the $K^-N$ 
subthreshold region, is the decisive mechanism that controls the self-consistent evaluation of 
corresponding $K^-$ optical potentials. While the two in-medium versions of the $K^-N$ scattering 
amplitudes yield considerably different potential depths Re$V_K$  at threshold, they give similar 
depths in the self-consistent calculations with the subthreshold extrapolation of $\sqrt{s}$.    
The role of hadron self-energies in the self-consistent calculations of the $K^-$ binding energies 
$B_K$ is less pronounced than the model dependence of predicted $B_K$ which amounts to 
$\Delta B_K \approx 15$~MeV. As for the calculated widths $\Gamma_K$, the model dependence 
and the effects due to the hadron self energies are comparable.
The p-wave interaction generated by the $\Sigma(1385)$ subthreshold resonance was found to play only a marginal role. 

The widths  of low-lying $K^-$ states due to $K^-N \rightarrow \pi Y$ conversions are substantially 
reduced in the self-consistent calculations, thus reflecting the proximity of the $\pi\Sigma$ threshold. 
On the contrary, the widths of higher excited $K^-$ states are quite large even if only the pion 
conversion modes on a single nucleon are considered. 
After including 2 body $K^-NN \rightarrow YN$ absorption modes, the total decay widths $\Gamma_K$ 
are comparable or even larger than the corresponding binding energies $B_K$ for \emph{all} 
$K^-$ nuclear quasi-bound states, exceeding considerably the level spacing. 

The above conclusions should discourage attempts to search for isolated peaks corresponding to $K^-$ 
nuclear quasi-bound states in any but very light nuclear systems.    

\section*{Acknowledgements}

We wish to acknowledge a fruitful collaboration with Ale\v{s} Ciepl\'{y}, Eli Friedman and Avraham Gal.  
This work was supported by the GACR grant No.~202/09/1441 and by the EU inciative FP7, HadronPhysics2, 
under project No.~227431. 
%
%

\end{document}